%% file: TTbarBoostedPosterProc.tex
\documentclass[12pt]{article}
\usepackage{graphicx}
\usepackage{xspace}
\usepackage{lineno}

\newcommand\pubnumber{
}
\newcommand\pubdate{\today}

\def\Title#1{\begin{center} {\Large #1 } \end{center}}
\def\Author#1{\begin{center}{ \sc #1} \end{center}}
\def\Address#1{\begin{center}{ \it #1} \end{center}}

\newcommand\pubblock{\rightline{\begin{tabular}{l} \pubnumber\\
         \pubdate  \end{tabular}}}
\newenvironment{Abstract}{\begin{quotation}  }{\end{quotation}}
\newenvironment{Presented}{\begin{quotation} \begin{center} 
             PRESENTED AT\end{center}\bigskip 
      \begin{center}\begin{large}}{\end{large}\end{center} \end{quotation}}

\input econfmacros.tex

\begin{document}

\newcommand{\gevn}      {\ensuremath{\mathrm{GeV}}}
\newcommand{\tev}      {\ensuremath{~\mathrm{TeV}}\xspace}
\newcommand{\gev}      {\ensuremath{~\mathrm{GeV}}\xspace}
\newcommand{\GeV}      {\ensuremath{~\mathrm{GeV}}\xspace}
\newcommand{\mev}      {\ensuremath{~\mathrm{MeV}}\xspace}
\newcommand{\pb}      {\ensuremath{~\mathrm{pb}}\xspace}
\newcommand{\m}      {\ensuremath{~\mathrm{m}}\xspace}
\newcommand{\cm}      {\ensuremath{~\mathrm{cm}}\xspace}
\newcommand{\ns}      {\ensuremath{~\mathrm{ns}}\xspace}

\newcommand{\ttbar}     {\ensuremath{t\bar{t}}\xspace}
\newcommand{\pt}        {\ensuremath{p_{\rm T}}\xspace}
\newcommand{\ptjet}        {\ensuremath{p_{\rm T}^{\mathrm{jet}}}\xspace}
\newcommand{\mtt}       {\ensuremath{m^{\ttbar}}\xspace}
\newcommand{\pttt}      {\ensuremath{\pt^{\ttbar}}\xspace}
\newcommand{\ptt}       {\ensuremath{\pt^{\rm{top}}}\xspace}
\newcommand{\ptl}    {\ensuremath{\pt^{t,1}}\xspace}
\newcommand{\pts}    {\ensuremath{\pt^{t,2}}\xspace}

\newcommand{\antikt}    {anti-\ensuremath{k_{t}}\xspace}
\newcommand{\akt}       {\antikt\xspace}
\newcommand{\largeR} {large-R\xspace}
\newcommand{\LargeR} {Large-R\xspace}

\begin{titlepage}
\pubblock

\vfill
\Title{Differential \ttbar cross-section measurements using boosted top quarks with 139 fb$^{-1}$ of ATLAS data}
\vfill
\Author{ Petr Ja\v{c}ka}
\Address{on
behalf of the ATLAS Collaboration}
\vfill
\begin{Abstract}
Preliminary results of two differential cross-section measurements in the ATLAS experiment at CERN are presented. Measurements are focused on the top-antitop quark pair production in the~lepton+jets and the~all-hadronic channels in the kinematic region with high transverse momentum top quarks. Both measurements use the full Run 2 data of 13~TeV proton-proton collisions from the Large Hadron Collider collected by the ATLAS detector in 2015-2018, corresponding to an integrated luminosity of 139~fb$^{-1}$. The measured spectra are compared with the Standard Model predictions. They are also used to set limits on the Wilson coefficients of an effective field theory extension of the Standard Model.

\end{Abstract}
\vfill
\begin{Presented}
$14^\mathrm{th}$ International Workshop on Top Quark Physics\\
(videoconference), 13--17 September, 2021 \\
\end{Presented}
\vfill
{\footnotesize Copyright 2022 CERN for the benefit of the ATLAS Collaboration. CC-BY-4.0 license.}
\end{titlepage}
\def\thefootnote{\fnsymbol{footnote}}
\setcounter{footnote}{0}

\section{Introduction}

This note summarizes preliminary results of two measurements of top-antitop (\ttbar) quark pair production differential cross-sections in the ATLAS experiment~\cite{PERF-2007-01}. Measurements are focused on high transverse momentum (\pt) top quarks in two channels: lepton+jets~\cite{ATLAS-CONF-2021-031} and all-hadronic~\cite{ATLAS-CONF-2021-050}. Both measurements use the full Run 2 dataset of 13~TeV proton-proton collisions from the Large Hadron Collider (LHC) collected by the ATLAS detector in 2015-2018, corresponding to an integrated luminosity of 139~fb$^{-1}$. The \ttbar production with high-\pt (boosted) top quarks is sensitive to deviations from the Standard Model (SM) prediction. This motivates for precise measurements in this topology. Results are compared with SM predictions and are used to set limits on the Wilson coefficients of an effective field theory (EFT) extension of the SM.

\section{Reconstruction of top-antitop pairs}
The two measurements use different strategies to reconstruct \ttbar pairs. From now on, top quarks decaying into one $b$-quark, one charged lepton and one neutrino are referred to as leptonic top quarks. Top quarks decaying into one $b$-quark and two additional quarks are referred to as hadronic top quarks.

The lepton+jets measurement is focused on events with one hadronic top and one leptonic top quark. The hadronic top quark is reconstructed by large-$R$ ($R=1.0$) \antikt jets, reclustered from $R=0.4$\ \antikt particle-flow jets. Large-$R$ jets are required to have $\pt>355$~GeV, $|\eta|<2.0$, and $|m - 170 \GeV|<50$~GeV, where $\eta$ is pseudorapidity and $m$ is the large-$R$ jet mass. In addition, the large-$R$ jet must contain at least one $R=0.4$~jet  identified as coming from the hadronization of $b$-quark ($b$-tagged). The leptonic top quark is reconstructed from a charged lepton (electron or muon), a $b$-tagged $R=0.4$ jet, and a reconstructed neutrino.

The all-hadronic measurement is focused on events with two hadronic top quarks. They are reconstructed using large-$R$  \antikt jets, clustered directly from calorimeter clusters. Jets are ordered according to \pt. The leading jet is required to have \\$\pt>500$\ GeV and the subleading jet is required to have $\pt>350$~GeV. Both jets are required to have $|\eta|<2.0$ and $|m-172.5 \GeV|<50$~GeV. In addition, both jets are required to be $b$-tagged and top-tagged.

The selection requirements above are used to define region of interest and to reduce background contributions. The most relevant sources of the background are listed in Table~\ref{tab:background}. Both measurements use data-driven techniques to determine multijet background originating mainly from strong interactions of the colliding protons. Contributions of other background sources are determined by Monte Carlo simulations normalised to theoretical predictions of their cross-sections. 

\begin{table}[!h!tbp]
\begin{center}
\caption{Main background processes ordered by their significance.}
\vspace{0.2cm}
\begin{tabular}{c|c} \hline
\textbf{Lepton+jets} & \textbf{All-hadronic} \\ \hline
Single-top & Multijet \\
\ttbar+$X(X=W,Z,H$) & \ttbar non-allhad \\
Multijet & Single-top \\
Others & \ttbar+$X(X=W,Z,H)$
\end{tabular}
\label{tab:background}
\vspace{-0.7cm}
\end{center}
\end{table}

\begin{figure}[!h!tbp]
\centering
\includegraphics[width=0.465\textwidth]{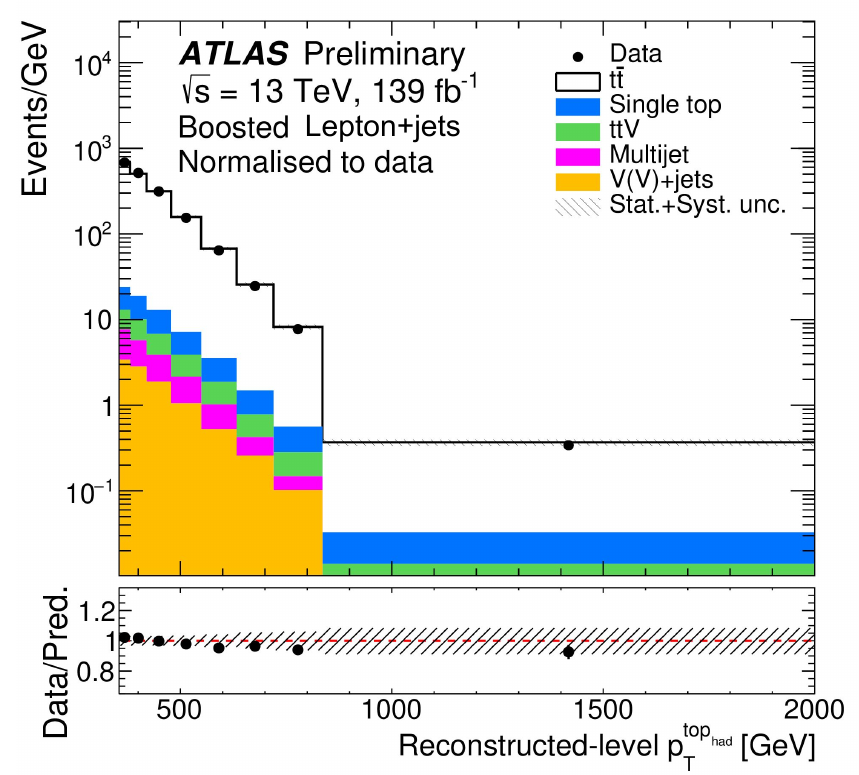}
\includegraphics[width=0.465\textwidth]{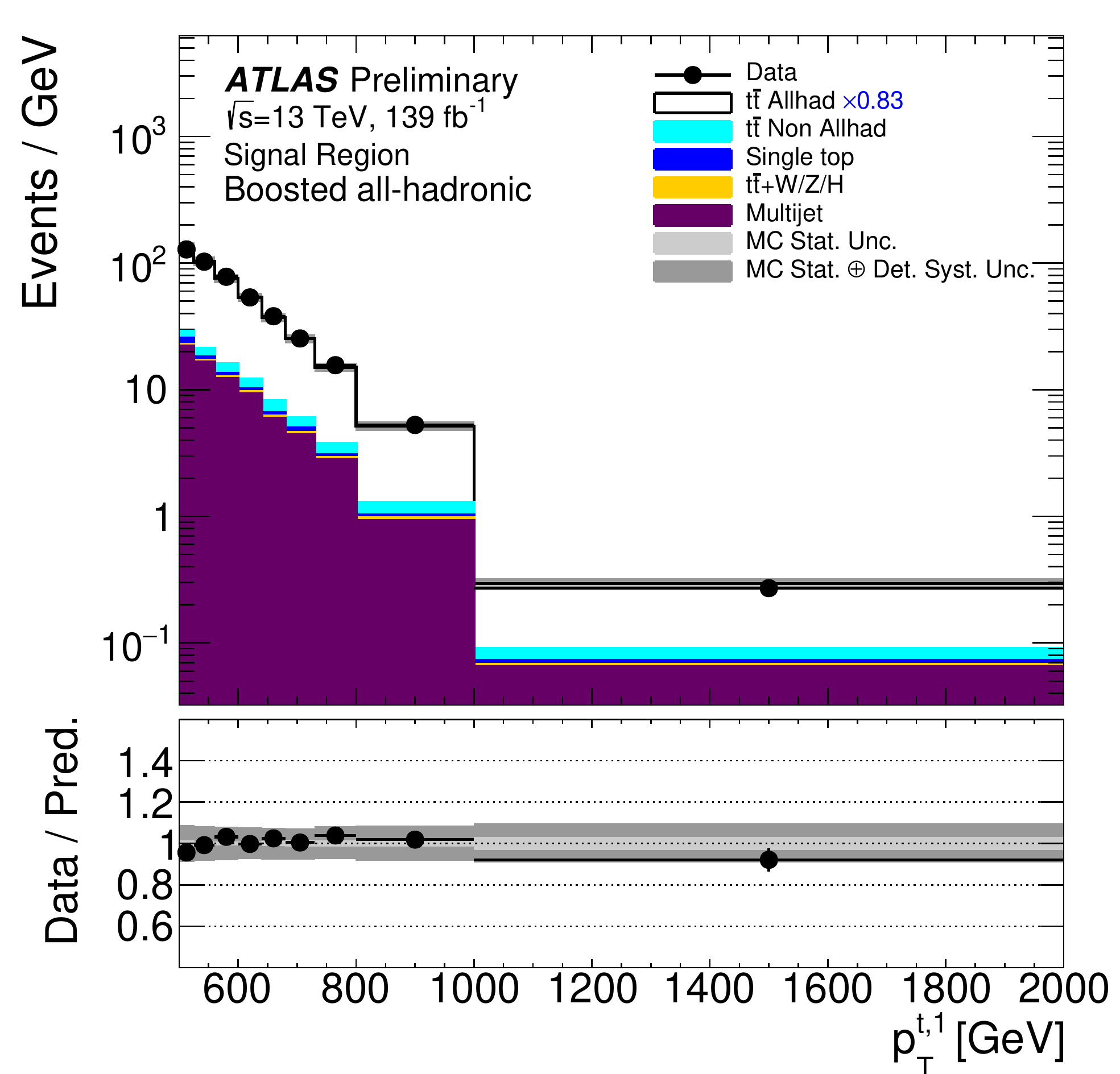}
\caption{Reconstructed hadronic top \pt in the lepton+jets channel (left)~\cite{ATLAS-CONF-2021-031} and leading top \pt in the all-hadronic channel (right)~\cite{ATLAS-CONF-2021-050}.  Predictions are normalized to match data normalization. The prediction is the sum of the signal and the background contributions.}
\label{fig:reco:pt}
\vspace{-0.3cm}
\end{figure}

Figure~\ref{fig:reco:pt} shows the reconstructed top \pt distributions in both channels. The spectra are steeply falling with \pt in both channels and both measurements are reaching 2~TeV. The data distributions are overlayed with the SM prediction, which is the sum of the signal and all background contributions. Predictions are normalized to match data normalization. Therefore, only the shapes of distributions can~be~compared.

\section{Unfolded results}
Reconstructed spectra are unfolded to a particle-level fiducial phase space designed to keep the measured phase-space as close as possible to that selected by the event selection. In addition, the all-hadronic analysis unfold spectra to a parton-level fiducial phase defined by requiring that the leading (subleading) top quark has\\$\pt>500 (350)\GeV$.

\begin{figure}[!h!tbp]
\vspace{-0.5cm}
\centering
\includegraphics[width=0.465\textwidth]{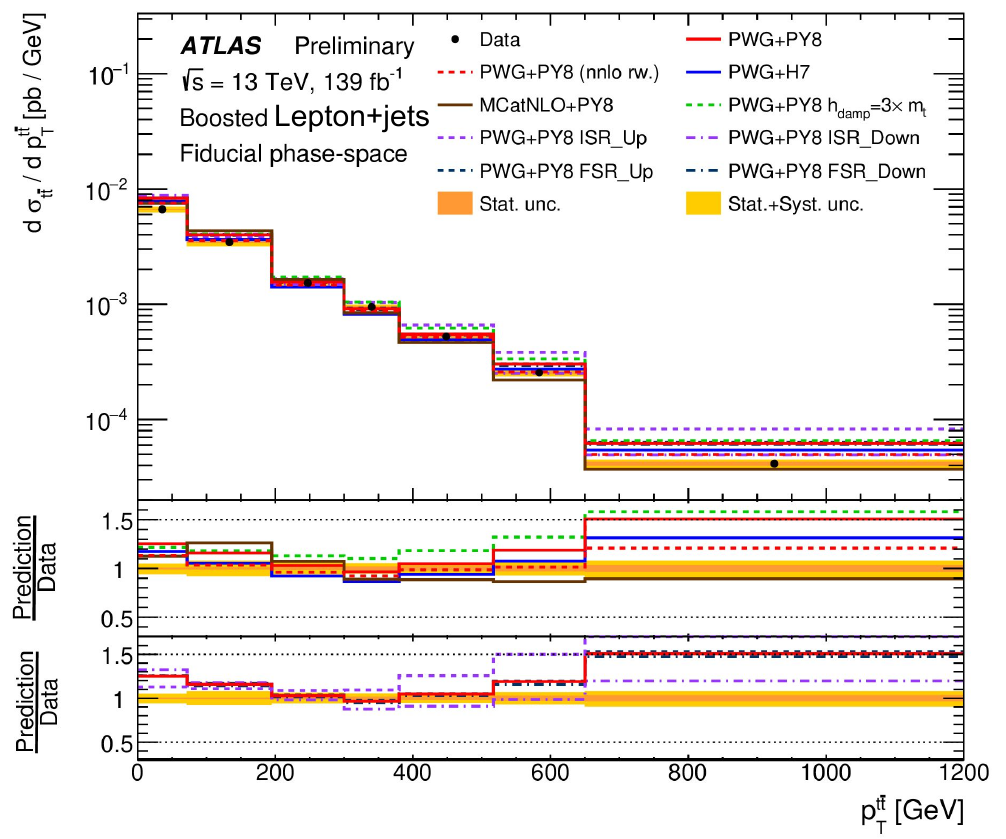}
\includegraphics[width=0.52\textwidth]{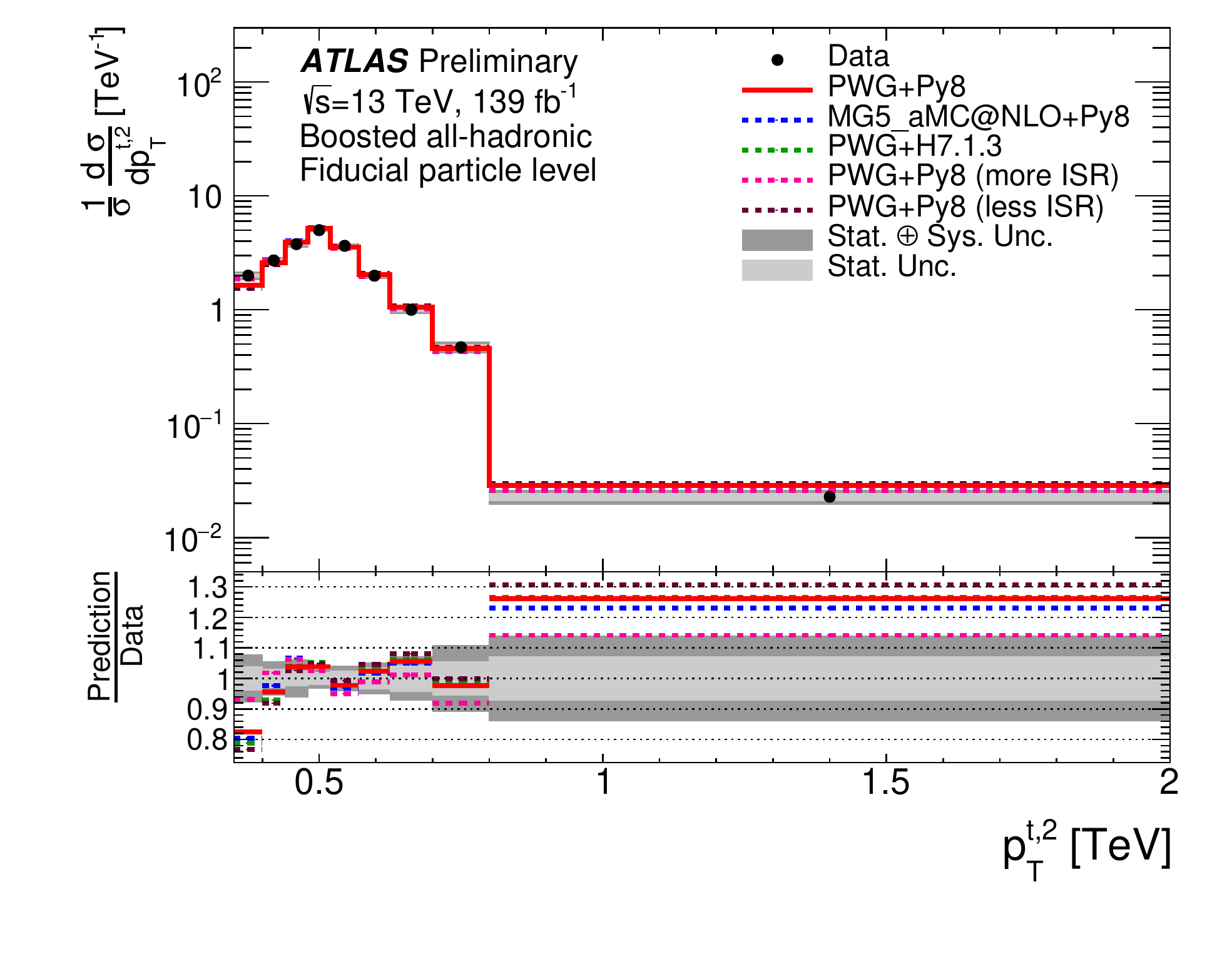}
\caption{\small{Particle-level fiducial phase-space differential cross-sections as a
function of hadronic top \pt in the lepton+jets channel (left)~\cite{ATLAS-CONF-2021-031}, and the \pt of the second-leading top-quark jet in the all-hadronic channel (right)~\cite{ATLAS-CONF-2021-050}. The distribution of the  the second-leading top-quark \pt is normalized to unity. 
Data points are placed in the center of each bin of measured spectra. The bands around data points indicate the total and the statistical uncertainty in each bin. The data are compared to multiple NLO+PS predictions represented by lines.}}
\label{fig:particle:pt}
\end{figure}

 Both analyses use a Bayesian iterative method with 2-4 iterations to regularize the unfolding problem. The all-hadronic measurements are further normalized to unity to cancel out systematic uncertainties in the normalization. 
Distributions at both levels are compared with \ttbar next-to-leading order + parton-showering (NLO+PS) generators: POWHEG+Pythia8, POWHEG+Herwig7, and MadGraph5 aMC@NLO+Pythia8 with various configurations. In addition, the parton level spectra are compared with the SM fixed order next-to-next-to leading order (NNLO) \ttbar differential cross-section prediction. Figure~\ref{fig:particle:pt} shows the measured particle-level differential cross-sections as a function of hadronic top \pt in the lepton+jets channel, and the \pt of the subleading top-quark jet in the all-hadronic channel. A small discrepancy is observed between data and predictions as well as a mild inconsistency between NLO+PS generators. 

\begin{figure}[!h!tbp]
\vspace{-0.5cm}
\centering
\includegraphics[width=0.49\textwidth]{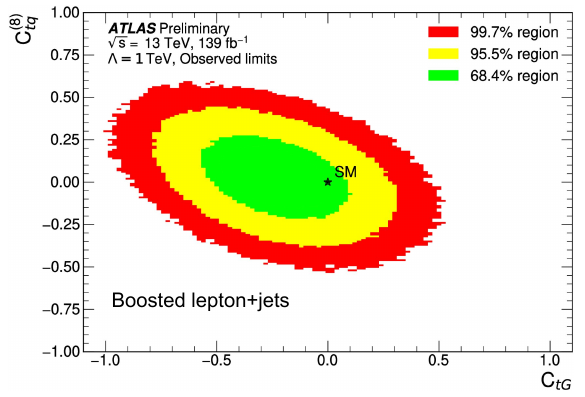}
\includegraphics[width=0.49\textwidth]{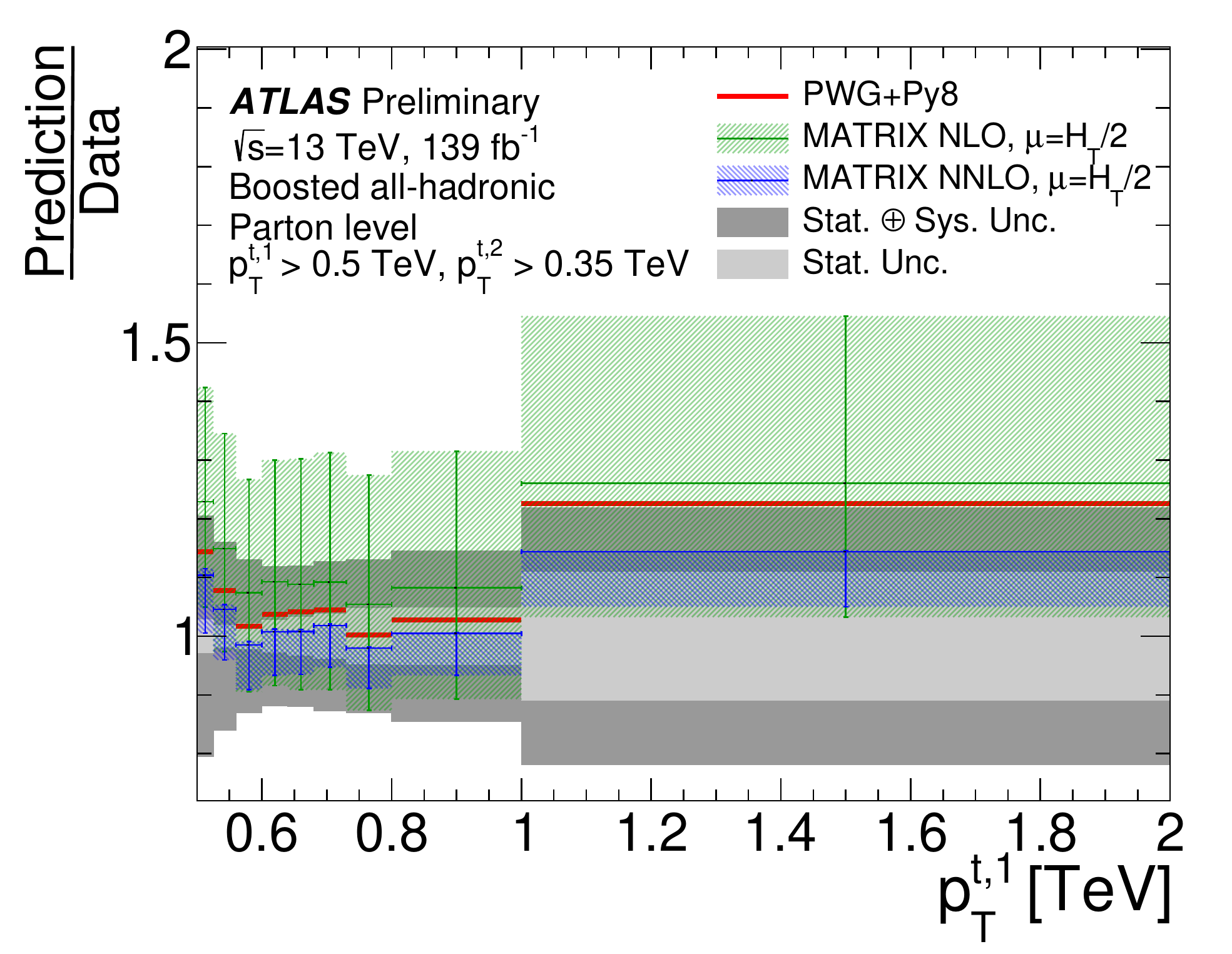}
\caption{\small{Two-dimensional limits on the EFT coefficients extending the Standard Model (left)\cite{ATLAS-CONF-2021-031} and comparison of the NLO, NNLO, and POWHEG+Pythia8 predictions with measured parton-level fiducial phase-space absolute differential cross-sections for ~the \pt of the
leading top-quark (right)~\cite{ATLAS-CONF-2021-050}.}}
\label{fig:parton:eft:pt}
\end{figure}

Figure \ref{fig:parton:eft:pt} (right) shows a comparison of the all-hadronic cross-section as a function of leading top quark transverse momentum with the SM fixed-order predictions and the POWHEG+Pythia8 predictions. All predictions are in good agreement with data. However, the NNLO prediction tends to be closer to data than other predictions. In addition, the NNLO prediction has smaller scale uncertainties with respect to the NLO prediction. This demonstrates a better precision and stability of higher-order predictions.

Measurements of top-quark differential cross-sections are further used to place constraints on Wilson coefficients $C_{i}$ in the EFT extension of the SM model defined~as
  $$\mathcal{L}_{\textrm{SM+EFT}} = \mathcal{L}_{\textrm{SM}} + C_{tG}\cdot\mathcal{O}_{tG}/\Lambda^2  + C_{tq}^{(8)}\cdot\mathcal{O}_{tq}^{(8)}/\Lambda^2,$$
where the $\mathcal{O}_{tG}$ and $\mathcal{O}_{tq}^{(8)}$ operators induce anomalous top-gluon and top-quark couplings and the energy scale $\Lambda$ is set to 1 TeV (Figure~\ref{fig:parton:eft:pt} left). The measured values of the coefficients are consistent with zero.


\section{Conclusions}
Preliminary results of two \ttbar differential cross-sections measurements in the ATLAS experiment at the LHC are presented. The measurements are focused on high-\pt top quarks in the lepton+jets and all-hadronic channels. Both measurements improved precision with respect to the previous measurements. In general, good agreement is observed between NLO+PS Monte Carlo simulations and the measured spectra. However, tensions are observed in distributions sensitive to radiations. In addition, measurements from the all-hadronic channel are compared with the fixed order NNLO predictions. NNLO predictions tend to be closer to the measured spectra with respect to NLO+PS predictions. Limits are set on selected coefficients of an effective field theory extending the Standard Model. Coefficients are found to be consistent~with~zero.


%
 

\end{document}

%% file: econfmacros.tex
\def\beq{\begin{equation}}
\def\eeq#1{\label{#1}\end{equation}}
\def\eeqn{\end{equation}}

\def\beqa{\begin{eqnarray}}
\def\eeqa#1{\label{#1}\end{eqnarray}}
\def\eeqan{\end{eqnarray}}

\let\bar=\overbar

\def\Dslash{\not{\hbox{\kern-4pt $D$}}}
\def\dslash{\not{\hbox{\kern-2pt $\del$}}}

\def\msb{{\bar{\ssstyle M \kern -1pt S}}}

